\documentclass[preprint,12pt]{elsarticle}
\usepackage{amssymb}
\usepackage{amsmath}

\usepackage{lineno}

\journal{Nuclear Instruments and Methods in Physics Research A}

\begin{document}

\begin{frontmatter}

\title{Modeling Charge Cloud Dynamics in Cross Strip Semiconductor Detectors}

\author[a,b]{Steven E. Boggs\corref{cor1}}
\ead{seboggs@ucsd.edu}
\cortext[cor1]{Corresponding author.}

\affiliation[a]{organization={Department of Astronomy \& Astrophysics, University of California, San Diego},
            addressline={9500 Gilman Drive}, 
            city={La Jolla},
            state={CA},
            postcode={92093}, 
            country={USA}}
            
\affiliation[b]{organization={Space Sciences Laboratory, University of California, Berkeley},
            addressline={7 Gauss Way}, 
            city={Berkeley},
            state={CA},
            postcode={94720}, 
            country={USA}}
\begin{abstract}

When a $\gamma$--ray interacts in a semiconductor detector, the resulting electron-hole charge clouds drift towards their respective electrodes for signal collection. These charge clouds will expand over time due to both thermal diffusion and mutual electrostatic repulsion. Solutions to the resulting charge profiles are well understood for the limiting cases accounting for only diffusion and only repulsion, but the general solution including both effects can only be solved numerically. Previous attempts to model these effects have taken into account the broadening of the charge profile due to both effects, but have simplified the shape of the profile by assuming Gaussian distributions. However, the detailed charge profile can have important impacts on charge sharing in multi-electrode strip detectors. In this work, we derive an analytical approximation to the general solution, including both diffusion and repulsion, that closely replicates both the width and the detailed shape of the charge profiles. This analytical solution simplifies the modeling of charge clouds in semiconductor strip detectors.

\end{abstract}



\end{frontmatter}


\section{Introduction}
\label{sect:intro}
\begin{figure}
\centering
\includegraphics[width=0.8\textwidth]{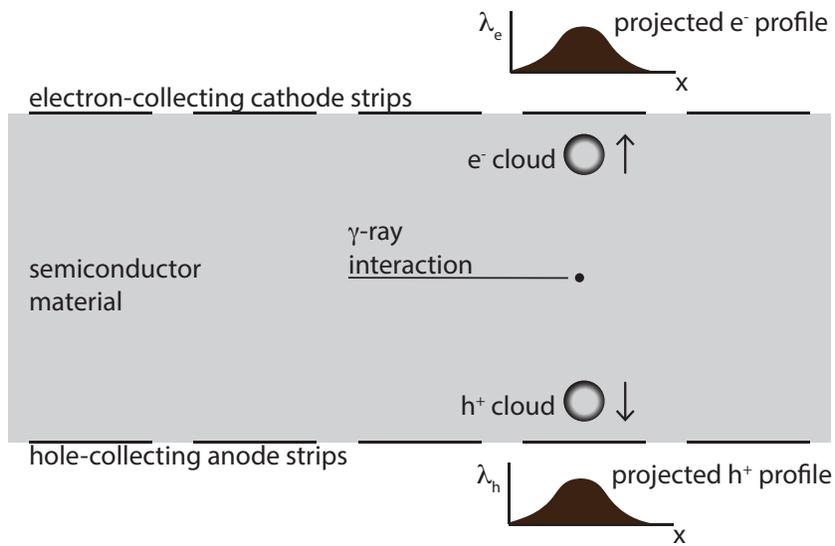}
\caption{\label{fig:f1} Diagram of the charge collection process in a cross-strip semiconductor detector induced by a $\gamma$--ray interaction. The resulting recoil electron will produce a cloud of electron-hole pairs that will drift, in an applied bias field, towards the cathode (electrons) and anode (holes) respectively. The charge clouds will broaden as they drift, potentially inducing signals on neighboring electrode strips in the process. We have depicted the strips on the cathode and anode non-orthogonally in this diagram for simplicity.}
\end{figure}

The development of large semiconductor detectors for $\gamma$--ray imaging and spectroscopy requires a detailed understanding of the charge cloud dynamics in order to optimize the measurement of both the location and the energy of the initial photon interaction. When a $\gamma$--ray photon interacts in a semiconductor detector, either by photoabsorption or Compton scattering, a fast recoil electron is produced which knocks more electrons from the valence band to the conduction band, leaving holes behind. The number of electron-hole (e-h) pairs, $N$, is directly proportional to the energy deposited, determined by the underlying band gap of the specific semiconductor. In an applied electric field these charge clouds will separate and drift in opposite directions, electrons toward the cathode and holes toward the anode (Fig.~\ref{fig:f1}). By electrically segmenting the cathode into strips, and the anode into orthogonal strips, 2-D positioning is achieved directly through identification of the electrode strips triggered for a given photon interaction. In thicker semiconductor detectors the difference between electron and hole collection times on opposite faces of the detector can be measured to determine the position of the $\gamma$--ray interaction between the cathode and anode strips.  Hence, full 3-D position resolution is achievable for localizing $\gamma$--ray interactions within the detector. 

In addition, the interaction energy can potentially be measured on both the cathode and the anode independently by measuring the integrated charge induced on each electrode by the charge clouds drifting towards their respective electrodes. As these charge clouds drift in the detector, their charge density profiles broaden due to both thermal diffusion and mutual electrostatic repulsion. The finite size of the charge clouds leads to some interactions having their charge collected on multiple electrodes. Such interactions result in either charge sharing between strips, or low-energy tailing on spectral lines if the charge shared on the neighboring strip falls below the detection threshold for that electrode. Charge sharing can also clearly affect the positioning performance of the detectors when multiple electrodes are triggered on one face of the detector. The detailed shape of these charge profiles can have important consequences for both low-energy tailing and charge sharing in multi-electrode strip detectors. Optimizing both the spectral and positioning performance of these detectors requires detailed knowledge of the broadened charge cloud profiles as they are collected on their respective electrodes.

We present a novel method of modeling the broadened charge cloud profiles projected in 1-D for utilization in strip detectors. The derived analytical formula closely approximates the full numerical solution to the charge continuity equation. Specifically, this method largely preserves the detailed profile of the broadened charge cloud, a significant improvement over previous approximations. Development of this analytical equation for the charge profile provides ease of application in detector performance modeling.

In Section~\ref{sect:cont-eqn}, we summarize limiting-case solutions of the charge cloud continuity equation that can help intuitively understand the full equation, which can only be solved numerically. In Section~\ref{sect:previous}, we discuss how the charge cloud dynamics have been approximated in previous works. Section~\ref{sect:convolution} presents our novel approximation method for simulating the extended charge cloud profiles. In Section~\ref{sect:extended}, we discuss how this model can be modified to account for an extended initial charge density. We conclude with a discussion of applications and future directions.

\section{Charge Cloud Dynamics}
\label{sect:cont-eqn}

In their classic paper, \citet{gatti1987dynamics} set up the full continuity equation for a spherical cloud of electrons drifting in silicon detectors, taking into account drift, diffusion, and electrostatic repulsion. They present solutions for the general case including the effects of both diffusion and repulsion which must be solved numerically. They also present analytical solutions for the two limiting cases: the diffusion-only case which dominates when the number of electrons in the charge cloud is relatively small, and the repulsion-only case which dominates when the number of electrons is relatively high. These solutions assume that an initial recoil electron loses all of its energy to e-h pairs at a single point in the detector. We revisit this assumption in Section~\ref{sect:extended}. These two limiting cases are useful for understanding and characterizing the physical dynamics of charge clouds in semiconductor drift detectors. 

In the diffusion-only solution (ignoring repulsion effects), the spherical cloud of $N$ charges follows a Gaussian profile with a characteristic radius $\sigma$ that expands as the square root of drift time:

\begin{equation}
\sigma (t) = \sqrt{2Dt}
\end{equation}

The diffusion constant, $D$, is determined by the mobility $\mu$ of the charge carrier and the temperature $T$ of the semiconductor by:

\begin{equation}
D=\mu \frac{kT}{e}
\end{equation}

Where $k$ is the Boltzmann constant and $e$ is the electron charge. We can project this spherical distribution along a single axis (x-axis) to represent the 1-D charge profile, $\lambda (x,t)$, collected across the electrode strip(s) of the detector:

\begin{equation}
\lambda (x,t) = \frac{N}{\sqrt{2 \pi} \sigma} e^{\frac{-x^2}{2\sigma^2}}
\label{eqn:diffusion}
\end{equation}

\begin{figure}
\centering
\includegraphics[width=1.0\textwidth]{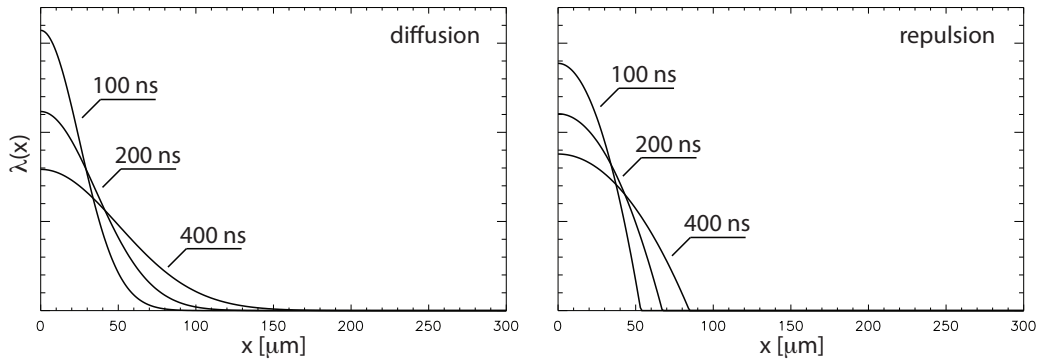}
\caption{\label{fig:f2} (Left) The diffusion-only spherical electron density distribution projected along the x-axis (Eqn.~\ref{eqn:diffusion}) for drift times of 100\,ns, 200\,ns, and 400\,ns. These curves assume a 122\,keV initial energy deposit in a room-temperature silicon detector. (Right) The repulsion-only solution (Eqn.~\ref{eqn:repulsion}) for the same detector assumptions.}
\end{figure}

Figure~\ref{fig:f2} (Left)  shows this diffusion-only 1-D solution in a room temperature silicon detector (assuming $\mu$ = 1250\,$cm^{2}/V \cdot s$) for a cloud of 33,700 electrons (122\,keV) at three drift times (100\,ns, 200\,ns, and 400\,ns).  

In the repulsion-only solution (ignoring diffusion effects), the spherical charge expands with a uniform charge density and a boundary radius $\eta$ that grows as the cube root of drift time:

\begin{equation}
\eta (t) = [\frac{3 \mu e}{4 \pi \epsilon} Nt]^{\frac{1}{3}}
\label{eqn:repulsion}
\end{equation}

Where $\mu$ is again the mobility of the charge carrier, and $\epsilon$ is the dielectric constant of the semiconductor material. The fact that $\eta$ scales as $N^{1/3}$ means that this repulsion distance increases with increasing number of mutually-repelling charge carriers in the cloud, becoming a significant effect compared to diffusion for higher interaction energies.

Again, we can project this spherical distribution along the single x-axis to represent the 1-D charge profile collected across the electrode strip(s):

\begin{equation}
\lambda (x,t) = \frac{3N}{4 \eta ^{3}}  [\eta ^2 - x^2]; |x| \le \eta (t)
\end{equation}

Figure~\ref{fig:f2} (Right) shows this repulsion-only 1-D solution again in a room temperature silicon detector for the same assumptions in Figure~\ref{fig:f2} (Left).

We can see that both diffusion and repulsion significantly contribute to the broadening of the charge cloud profile in room-temperature silicon detectors at these interaction energies. Hence it is important that we model both effects well in order to understand and optimize the performance of any $\gamma$--ray semiconductor detector.

\section{Previous Approximations}
\label{sect:previous}

As mentioned in Section~\ref{sect:cont-eqn}, the full continuity equation including the effects of both diffusion and repulsion can only be solved numerically. Given the lack of an analytical solution, researchers have adopted various approaches to estimate the extent of the broadened charge clouds and the resulting effects on charge collection in all types of semiconductor detectors. 

The simplest approach has been to justify that repulsion does not play a significant role in their particular detector application or else to ignore the repulsion effects altogether \cite{du1999monte,he2000effects,kim2011charge}. In these cases, the exact diffusion-only charge distributions are assumed adequate to model the effects of broadening on the charge collection. As noted above, such a solution can only be assumed for relatively low $\gamma$--ray interaction energies. 

In their original paper, \citet{gatti1987dynamics} derive the RMS width of the exact 1-D projected spherical densities as a function of drift time for a range of numbers of electrons in the charge cloud. Another approach is to utilize these RMS values directly, but treat them as the characteristic length scale of a Gaussian distribution \cite{bellwied2000studies}. This approach presumably better reflects the effects of repulsion despite the fact that the actual numerical solutions have more complicated profiles than simple Gaussians.

Another approach has been to take the RMS values of the separate diffusion-only and repulsion-only limiting solutions, and add them in quadrature to create a combined RMS which is treated as the characteristic radius in a Gaussian distribution \cite{donmez2005continued,nakhostin2015influence}. This approach has the advantage of being easy to implement analytically and accounting for both the diffusion and repulsion effects. However, this approach has two limitations. First, by assuming a Gaussian profile this approach does not reliably reproduce the charge profile for clouds with relatively large numbers of charges (higher energy events). Second, this approach overestimates the width of the resulting charge profiles, which \citet{benoit2009simulation} attribute to the method ignoring the mutual effects of these processes. We revisit this effect in Section~\ref{sect:disc}.

A more advanced approach has been to define an effective diffusion coefficient, $D’$, which includes the effects of both diffusion and repulsion \cite{benoit2009simulation}. This approach essentially starts with the pure diffusion solution, then utilizes that solution within the continuity equation to derive an effective time-dependent diffusion coefficient that includes the effects of repulsion. This effective diffusion coefficient decreases over time as the distribution increases in size, eventually approaching the purely thermal diffusion coefficient. This approach is intended to physically couple the diffusion and repulsion effects to better estimate the characteristic RMS widths at higher energies. However, this approach is still limited by assuming a Gaussian charge profile. Regardless of this limitation, this advanced approached has been adopted by multiple authors for studying the effects of charge cloud dynamics on charge collection in position-sensitive semiconductor detectors, e.g., \cite{altingun2022optimization,tang2021cadmium}. 

A similarly advanced approach was developed in \citet{bolotnikov2007cumulative}. In this approach, the solution to the full continuity equation including the effects of both diffusion and repulsion was approximated through a numerical algorithm developed by the authors that models the electron cloud as a spherically-symmetric set of infinitely thin shells. For a given initial charge distribution, the algorithm utilizes approximations to the continuity equation to trace the evolution of the shell radii with time. This approach is similarly intended to better couple the effects of both diffusion and repulsion to calculate a characteristic RMS width of the charge cloud as a function of drift time, which the authors then parameterize through a polynomial fit to the simulated results. This approach has the advantage of utilizing the full continuity equation to trace the evolution of the charge cloud, but is again ultimately limited by utilizing the characteristic RMS width value assuming a Gaussian charge profile.  

\section{Convolved Diffusion-Repulsion Approximation}
\label{sect:convolution}

Given the importance of both diffusion and repulsion for photon interactions at $\gamma$--ray energies in our cross-strip germanium detectors, we are motivated to find another approach to approximating the charge cloud distribution that not only accounts for both diffusion and repulsion on the width of the distribution, but which also better approximates the complex shape of the exact numerical solutions when the repulsion effects are significant. In this work, we are also motivated to find an analytical solution for ease of implementation in charge collection simulations. 

While the full solution of the continuity equation can only be solved numerically, we hypothesize that it may be approximated by convolving the diffusion-only solution and the repulsion-only solution. Comparison to the full solution will determine whether this hypothesis is justified. Intuitively, we can think of this approach as adopting the repulsion-only solution as the baseline and then broadening that solution by thermal-diffusion of the charges through the convolution process. 

\begin{equation}
\lambda (x,t) = \frac{3N}{4 \eta ^{3}}  \frac{1}{\sqrt{2 \pi} \sigma} \int_{x - \eta}^{x + \eta} e^{\frac{-t^2}{2\sigma^2}} [\eta ^2 - (x-t)^2] \,dt
\end{equation}

Fortunately, though complicated, this convolution has an analytical solution, making it a good candidate for the 1-D charge distribution approximation:

\begin{equation}
\label{eqn:convolution}
\begin{aligned}
\lambda (x,t) &= \frac{3N}{8 \eta ^{3}}  (\eta ^2 -x^2 - \sigma ^2) [erf(\frac{b}{\sqrt{2} \sigma}) -erf(\frac{a}{\sqrt{2} \sigma})]  \\ 
&\quad + \frac{3N}{4 \eta ^{3}} \frac{\sigma}{\sqrt{2 \pi}}  [(2x-a) e^ {\frac{-a^2}{2 \sigma ^2}} - (2x-b) e^ {\frac{-b^2}{2 \sigma ^2}} ]; \\ 
&\quad a \equiv x - \eta, b \equiv x + \eta
\end{aligned}
\end{equation}

The test of this charge profile distribution is how well it can reproduce the full numerical solutions of the continuity equation. 

\begin{figure}
\centering
\includegraphics[width=0.6\textwidth]{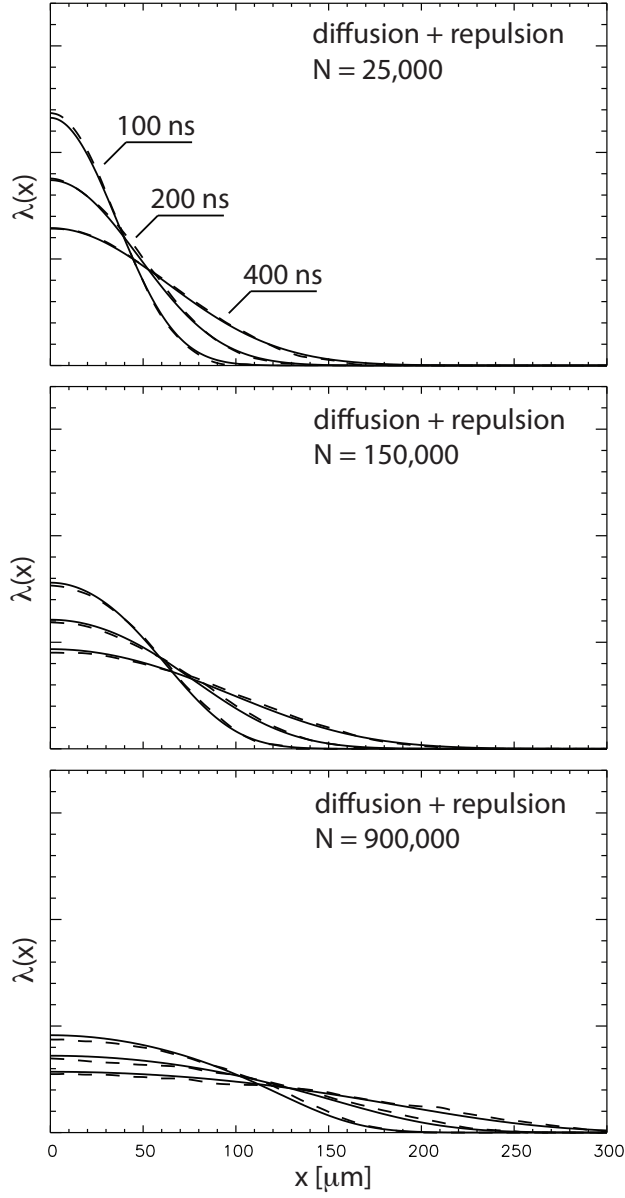}
\caption{\label{fig:f3} (Top) Comparison between the exact numerical continuity equation solutions in \citet{gatti1987dynamics} (dashed lines) compared with our analytical diffusion-repulsion approximation (Eqn.~\ref{eqn:convolution}, solid lines) at three drift times. This is for a cloud of 25,000 drift electrons, corresponding to an initial energy deposit of 90\,keV. (Middle) For 150,000 electrons (540\,keV). (Bottom) For 900,000 electrons (3300\,keV).}
\end{figure}

In Figure~\ref{fig:f3}, we compare our 1-D projected density profiles for a spherical cloud of electrons drifting in a room-temperature silicon detector with the exact numerical distributions presented in \citet{gatti1987dynamics}. The primary uncertainty in this comparison is the value of the electron mobility assumed in the original paper. While the assumed value is not explicitly given in the paper, the discussion presented at the end of the manuscript indicates that the value adopted for the original solutions was $\sim$1250\,$cm^{2}/V \cdot s$. This value is low compared to the electron mobility that we would have assumed a priori (1350\,$cm^{2}/V \cdot s$), but we have adopted the lower value for the profiles shown in Figure~\ref{fig:f3} to allow direct comparison with the original solutions. 

As is evident in these plots, our convolved solution does an excellent job in reproducing the effects of repulsion on both the width of the distribution as well as the profile of the distribution. Though not exact, Eqn.~\ref{eqn:convolution} comes much closer to replicating the exact solution including diffusion and repulsion effects than if we had simply modeled the distribution as a Gaussian. The details of this shape can have important consequences for both low-energy tailing and charge sharing in multi-electrode strip detectors. 

While the original diffusion-only and repulsion-only solutions were derived specifically for electron clouds in room-temperature silicon detectors, there is nothing unique in these solutions to the charge carrier (electrons, holes), the semiconductor detector material (Si, Ge, CdZnTe), or operating temperature (room temperature, cryogenic). The same holds for our convolved solution. We just need to adopt the correct charge mobility (unique to the carrier, material, and temperature), material dielectric properties, and operating temperature in defining the diffusion coefficient and the repulsion size scale.

\begin{figure}
\centering
\includegraphics[width=1.0\textwidth]{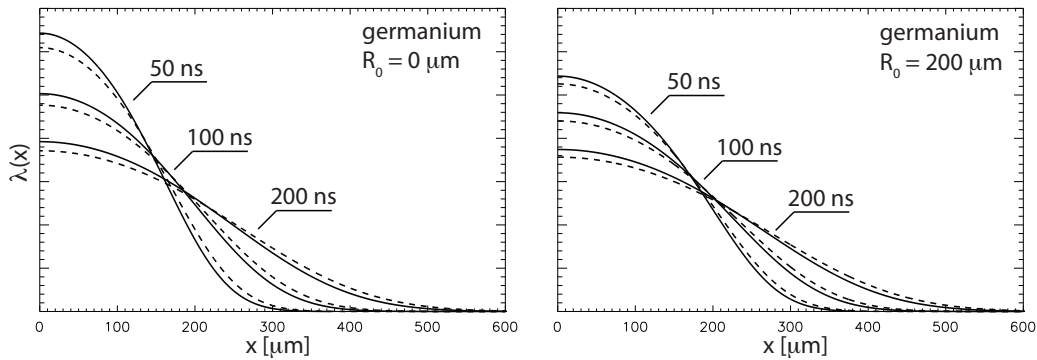}
\caption{\label{fig:f4} (Left) Our analytical diffusion-repulsion approximation (Eqn.~\ref{eqn:convolution}) applied to cryogenic germanium detectors for an initial 662\,keV $\gamma$--ray interactions. Electrons (solid lines) and holes (dashed lines) charge profiles are shown at three drift times. (Right) The same detector assumptions, but with an initial uniform charge cloud radius of $R_0$ = 200\,$\mu$m.}
\end{figure}

For example, in the large-volume germanium detectors we utilize for $\gamma$--ray astrophysics (described below) we collect both the electron signals and the holes signals on opposite faces of the detector. These detectors operate at cryogenic temperatures (80\,K) for optimal spectral resolution. Spreading of the charge clouds across neighboring strips can result in both low-energy spectral tailing and charge sharing between neighboring electrodes, which we need to understand in detail to optimize the performance of these detectors. In Figure~\ref{fig:f4} (Left) we show the expected charge profiles for an initial 662\,keV interaction at several drift times (50\,ns, 100\,ns, 200\,ns) for both the electrons ($\mu _e$ = $3.6 \times 10^4$ \,$cm^{2}/V \cdot s$) and holes ($\mu _h$ = $4.2 \times 10^4$ \,$cm^{2}/V \cdot s$). Having such detailed charge profiles will enable us to understand the effects of low-energy tailing and charge sharing in our detectors in much greater detail than previous simulations. 

\section{Extended Initial Charge Cloud}
\label{sect:extended}

There are two physical interactions that we have ignored in the analysis so far. The first is the detailed electron-hole plasma effects immediately after generation of the e-h charge clouds \cite{tove1967plasma}. These effects dissipate quickly compared to the drift timescales we are considering for charge collection, and hence are beyond the scope of this work. 

The second physical complication that we have ignored until this point is the impact of the extended distribution of electron-hole pairs created by the recoil electron following the initial $\gamma$--ray interaction. The solutions presented above assume that the $\gamma$--ray interaction creates an initial electron-hole cloud at a single point in the detector. In reality, the initial charge distribution is quite complicated depending on the detailed path followed by the recoil electron as it dissipates its energy through the creation of electron-hole pairs. Extended initial charge cloud distributions will become increasingly important for higher interaction energies, such as those we are measuring in $\gamma$--ray applications of our germanium detectors. 

The simplest approach to modeling the effects of an initially extended charge cloud in light of the solutions presented above is to assume that the initial charge cloud can be approximated at $t=0$ as a sphere of uniform charge density and finite radius $R_0$. Such a solution could adopt an average $R_0$ for a given interaction energy based on either the practical range for electrons in the semiconductor \cite{wohl1984review}, or else on detailed Monte Carlo models of ensembles of interactions. Likewise, we can estimate a specific $R_0$ for each interaction based on detailed Monte Carlo modeling of individual interactions. Regardless of the source of $R_0$, incorporating this initial condition into the solutions presented above is a simple matter of adjusting the repulsion-only solution to account for starting from a finite $R_0$ instead of a single point. This is effectively shifting the repulsion-only solution forward in time to the point of reaching $R_0$. This solution results in redefining the characteristic size scale for the repulsion-only solution to be:

\begin{equation}
\eta (t) = [R_0 ^{3} + \frac{3 \mu e}{4 \pi \epsilon} Nt]^{\frac{1}{3}}
\end{equation} 

Now that we have defined this refined repulsion-only size scale to account for the initial charge cloud size, we can utilize this revised $\eta$ directly in the convolved diffusion/repulsion solution in Eqn.~\ref{eqn:convolution}. 

For example, in germanium the practical range for 662\,keV electrons is roughly 400\,$\mu$m. Adopting $R_0$ = 200\,$\mu$m, in Figure~\ref{fig:f4} (Right) we show the effects of this initially extended charge cloud on our convolved solutions. 

There is another potential approach to modeling extended initial charge clouds that could allow more flexibility in the initial charge profile. \citet{benoit2009simulation} derived a solution to the repulsion-only continuity equation for an elliptical initial charge cloud of uniform charge density. Such an elliptical solution allows the possibility of modeling more complicated initial charge cloud distributions than the purely spherical approach discussed here. In principle, this elliptical solution can be convolved with the diffusion-only solution to create an analytical approximation for this more complicated scenario. Those calculations are beyond the scope of this work.

\section{Discussion}
\label {sect:disc}

In the model where RMS values are independently derived for each solution then added in quadrature to achieve an effective RMS \cite{donmez2005continued,nakhostin2015influence}, the resulting Gaussian profile tends to overestimate the width of the resulting charge clouds, which \citet{benoit2009simulation} attribute to this method ignoring the mutual effects of repulsion and diffusion. Given the agreement between our convolved diffusion-repulsion approximation and the exact numerical solutions from \citet{gatti1987dynamics}, the convolution technique presented here does not significantly overestimate the width of the charge distributions even though we do not explicitly correct for any mutual effects of repulsion and diffusion. Our result suggests that the overestimates in the effective RMS technique are due to the assumption of a Gaussian profile for the repulsion effects and are not an artifact of mutual interactions.

While we have developed this charge cloud approximation for application in any $\gamma$--ray semiconductor strip detector, we are particularly motivated to optimize the performance of the Compton Spectrometer and Imager (COSI). COSI is a soft $\gamma$--ray survey telescope (0.2-5\,MeV) designed to probe the origins of Galactic positrons, reveal sites of ongoing element formation in the Galaxy, use $\gamma$--ray polarimetry to gain insight into extreme environments, and explore the physics of multi-messenger events \cite{kierans20172016,sleator2019benchmarking,arxiv.2109.10403}. The COSI detectors are custom, large-volume (54\,cm$^2$ area, 1.5\,cm thick) cross-strip germanium detectors utilizing amorphous contact technologies \cite{amman2007amorphous}. The approximate analytical charge profile solution (Eqn.~\ref{eqn:convolution}) allows us to study the effects of charge cloud broadening on localization and resulting spectral profiles in detail, providing a powerful tool for understanding and optimizing our overall detector performance. Initial application of this approximation to our 2-mm pitch strip COSI detectors reproduces the profile and amplitude of our measured low-energy tails on photopeak spectral lines very well. Detailed applications of this approximation with the COSI detectors will be the subject of future work.

\section{Acknowledgements}
\label{}

This work was supported by the NASA Astrophysics Research and Analysis (APRA) program, grant 80NSSC21K1815. Thanks to A. Shih, J. Tomsick, and J. Roberts for useful discussions.

\bibliographystyle{elsarticle-num-names} 
\bibliography{refs}

\end{document}